\documentclass[twocolumn,preprintnumbers,superscriptaddress]{revtex4}
\usepackage{amsmath}
\usepackage{dcolumn}
\usepackage{bm}
\usepackage{epsfig}
\usepackage{epstopdf}
\usepackage{graphicx}
\usepackage{amsfonts}
\usepackage{amssymb}
\usepackage{mathrsfs}
\usepackage{color}
\usepackage{multirow}
\usepackage{appendix}

\begin{document}
\title{Ion-crystal demonstration of structural phase transition induced solely by temperature}
\author{J. Li}
\affiliation{State Key Laboratory of Magnetic Resonance and Atomic and Molecular Physics,
Wuhan Institute of Physics and Mathematics, Chinese Academy of Sciences, Wuhan, 430071, China}
\affiliation{School of Physics, University of the Chinese Academy of Sciences, Beijing 100049, China}
\author{L. L. Yan}
\email{qingnuanbinghe@126.com}
\affiliation{State Key Laboratory of Magnetic Resonance and Atomic and Molecular Physics,
Wuhan Institute of Physics and Mathematics, Chinese Academy of Sciences, Wuhan, 430071, China}
\author{L. Chen}
\email{liangchen@wipm.ac.cn}
\affiliation{State Key Laboratory of Magnetic Resonance and Atomic and Molecular Physics,
Wuhan Institute of Physics and Mathematics, Chinese Academy of Sciences, Wuhan, 430071, China}
\author{Z. C. Liu}
\affiliation{State Key Laboratory of Magnetic Resonance and Atomic and Molecular Physics,
Wuhan Institute of Physics and Mathematics, Chinese Academy of Sciences, Wuhan, 430071, China}
\affiliation{School of Physics, University of the Chinese Academy of Sciences, Beijing 100049, China}
\author{F. Zhou}
\affiliation{State Key Laboratory of Magnetic Resonance and Atomic and Molecular Physics,
Wuhan Institute of Physics and Mathematics, Chinese Academy of Sciences, Wuhan, 430071, China}
\author{J. Q. Zhang}
\affiliation{State Key Laboratory of Magnetic Resonance and Atomic and Molecular Physics,
Wuhan Institute of Physics and Mathematics, Chinese Academy of Sciences, Wuhan, 430071, China}
\author{W. L. Yang}
\affiliation{State Key Laboratory of Magnetic Resonance and Atomic and Molecular Physics,
Wuhan Institute of Physics and Mathematics, Chinese Academy of Sciences, Wuhan, 430071, China}
\author{M. Feng}
\email{mangfeng@wipm.ac.cn}
\affiliation{State Key Laboratory of Magnetic Resonance and Atomic and Molecular Physics,
Wuhan Institute of Physics and Mathematics, Chinese Academy of Sciences, Wuhan, 430071, China}
\affiliation{Center for Cold Atom Physics, Chinese Academy of Sciences, Wuhan 430071, China}
\affiliation{Department of Physics, Zhejiang Normal University, Jinhua 321004, China}

\begin{abstract}
We demonstrate for the first time a linear-zigzag phase transition induced solely by temperature of the $^{40}$Ca$^{+}$ ion crystals in a surface-electrode trap. In contrast to the previously observed counterparts based on change of the mechanical equilibrium conditions of the ions, our presented structural phase transition occurs due to controllable influence of thermal fluctuation. The ions' temperature is well controlled by tuning the cooling laser and the experimental observation could be fully understood by classical Langevin equation in addition to the effects from thermal fluctuation. Our experimental investigation indicates the fantastic role of thermal fluctuation played in the thermodynamic process at atomic level, which might bridge the thermodynamics from the macroscopic domain to the quantum regime.
\end{abstract}
%\pacs{05.70.-a,37.10.Vz,03.67.-a}
\maketitle

With advances of laser cooling techniques, observing thermodynamic process in few microscopic particles in cryogenic situation is currently available. In contrast to
the conventional thermodynamics applied in macroscopic systems, microscopic thermodynamics has displayed marvellous differences, ranging from violation of the basic thermodynamic laws to
the redefinition of thermodynamic quantities. Quantum thermodynamics is a typical development in this aspect \cite{parrondo}.

In the present Letter, we demonstrate an experiment relevant to configurational variation of cold ion crystals confined in an electromagnetic potential, which belongs to cryogenic phase transition regarding self-organized matter \cite{crystal2,crystal3,crystal4}. The critical behavior in such phase transitions occurred usually due to variation of characteristic parameters of the systems, which could be fully explained by classical physics \cite{review1}. These observations, including complex changes of the crystalline configuration \cite{config1,config2,config3,config6,config7,config8}, reflect the roles of nonlinearity and long-range order played in many-body physics. Understanding these characteristics is one of the indispensable prerequisites to the control of dynamics of the complex system at the microscopic level.

Dynamics of microscopic particles is usually subject to thermal fluctuation, which is generally regarded as a detrimental effect. But here we achieve a temperature-induced structural phase transition (SPT) in an ion-crystal system, in which thermal fluctuation could act positively. The SPT is a transformation between the linear structure to the zigzag, belonging to the configurational variation mentioned above \cite{config1,config2,config3,config6,config7,config8}, which was conventionally achieved by changing mechanical equilibrium conditions of the ions. However, the crystalline configuration in our case varies due to controllable change of thermal fluctuation experienced by the ions. Since this fluctuation effect is very feeble and hard to precisely control, generating such a SPT has never been explored before.
In fact, this fluctuation effect has also been neglected in previously theoretical studies focusing only on mechanical effects in the laser-ion interaction and Coulomb repulsion, and thus the experimental investigations in this aspect were based on the change of mechanical conditions, such as the trapping potential, e.g., our previous observation \cite{sr-6-21547}. Actually, the thermal fluctuation (even quantum fluctuation) effect, although very weak,
is another possible factor which could vary the structures of the ion crystals, as predicted in \cite {prl-105-265703}. Since this fluctuation effect can be reflected by the ions' temperature, the associated SPT is called temperature-driven SPT, which, in terms of the calculation in \cite {prl-105-265703}, occurs at the ions' temperature ranging from 1 mK to tens of mK.

\begin{figure}[htbp]
\centering
\includegraphics[width=8.5 cm,height=5.6 cm]{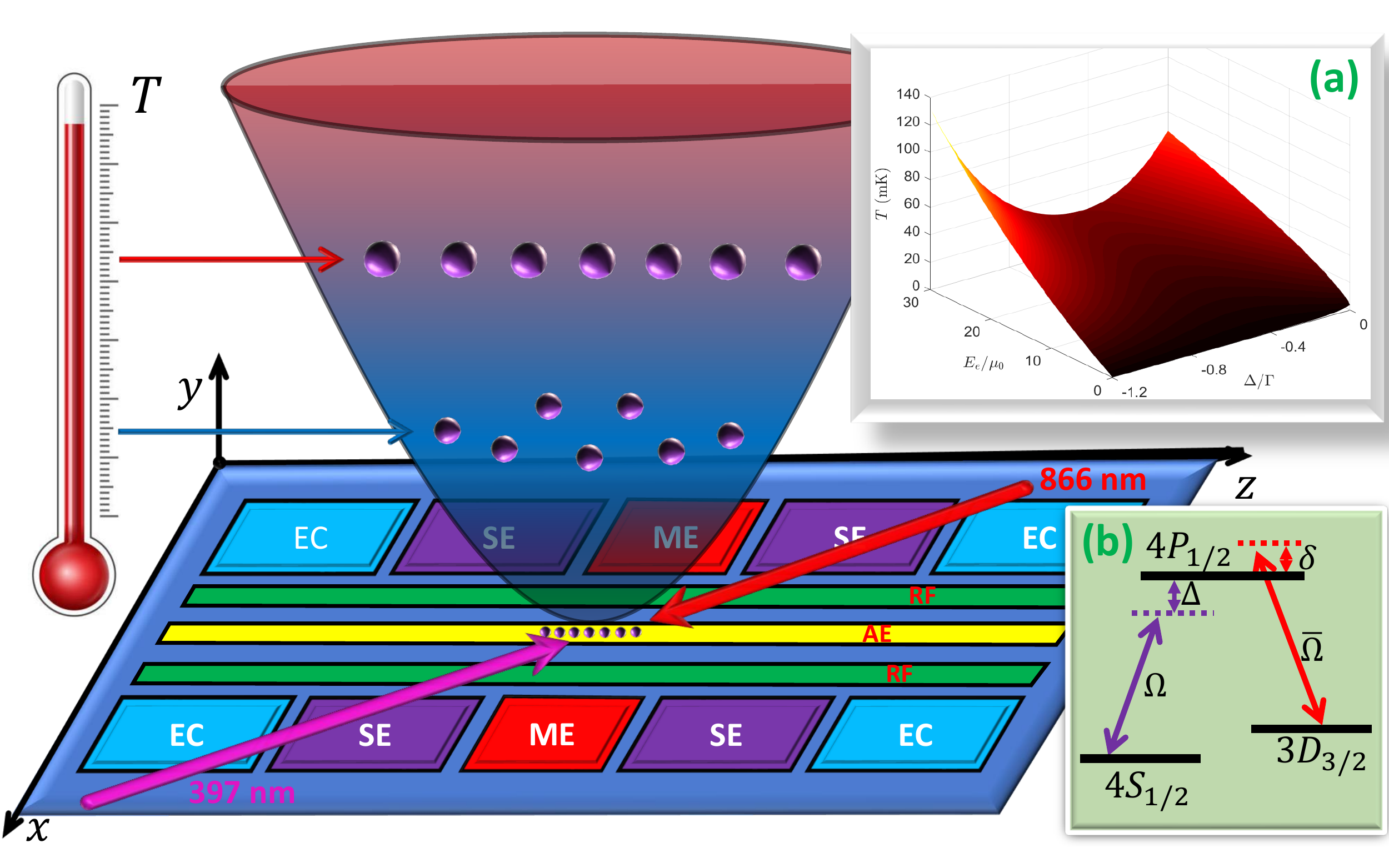}
\caption{Schematic of the surface-electrode trap and the SPT. The trap is composed of a central electrode (AE), two rf electrodes (RF) and two outer segmented dc electrodes. Each of the outer segmented dc electrode consists of five component electrodes, i.e., a middle electrode (ME), two control electrodes (SE) and two end electrodes (EC).
%The AE, RF and the gaps in between are of the same width of 500 $\mu$m. The widths of the SE and the EC are 1.5 mm, and the ME is 1 mm wide.
The ion crystals are irradiated by two lasers and our observation is along $y$-axis. The paraboloid above the trap plane presents a magnified diagram of a harmonic potential for seven crystalline ions varying structure with respect to the temperature $T$. Inset: (a) Temperature $T$ of the ion crystals relying on both the detuning $\Delta$ of the cooling laser and the bath heating intensity $E_{e}$ in units of $\mu_0=1\times 10^{-21} $ N$^2$s$/$kg, and (b) 397-nm laser drives the transition between the excited state $|4P_{1/2}\rangle$ (lifetime $7.1$ ns) and the ground state $|4S_{1/2}\rangle$, which plays the main role of cooling, and 866-nm laser couples $|4P_{1/2}\rangle$ to the metastable state $|3D_{3/2}\rangle$ for repumping the 6$\%$ leakage of spontaneous emission back to $|4P_{1/2}\rangle$. The detunings $\Delta$ and $\delta$ are defined as the frequency differences of the lasers from the two-level resonance.}
\label{Fig1}
\end{figure}

%Most of the recent observations of cryogenic phase transitions are relevant to quantum phase transitions \cite{Qexp}, which happen at zero temperature by changing some parameters in the systems regarding the properties of the ground state or the structure of the excited states \cite{Sachdev}. For some cryogenic phase transitions regarding self-organized matter, such as Wigner crystals of trapped ions in electromagnetic potentials \cite{crystal2,crystal3,crystal4},
%In particular, for the trapped-ions themselves, the various conformations of the ion crystals help for clarifying the influence from the rf heating - a detrimental noise against the ions' trapping. Since the temperature of the ions is influenced by the rf heating, the phase transitions regarding the ion crystal conformation are plausible to have relevance to temperature.

Specifically, our observation focuses on the linear-zigzag variation of seven laser-cooled $^{40}$Ca$^{+}$ ions, due to temperature change, in a surface-electrode trap, as sketched in Fig. \ref{Fig1}.
The trap is a 500-$\mu$m scale planar trap with five electrodes made of copper on a vacuum-compatible printed circuit board substrate. The details of the trap potential could be found in \cite{sr-6-21547}. One particular point worths mentioning here is the strong asymmetry in the trap, where the potential well in the $y$-axis is much steeper than in both $x$- and $z$-axes, leading to the ion crystals distributed only in the $xz$ plane. So the anisotropic parameter $\alpha=\omega_{x}/\omega_{z}$, with $\omega_{x}$($\omega_{z}$) being the trap frequency in $x$($z$)-axis, is an important quantity, whose variation over the transition point $\alpha_{c}$ drives the system to across from one phase to the other \cite{sr-6-21547}. In our case here, under the condition of $\alpha>3.14$, the ion crystals behave as one (line) or two (zigzag)-dimensional configurations in $xz$ plane with the thickness less than 2 $\mu$m along $y$-axis. As a result, our observation is along $y$-axis, which could obtain a full information about the structural changes of the ion crystals.

For our purpose, the key point of our implementation is the capability to controllably adjust the ions' temperature, which is relevant to the friction due to the cooling laser in this work. As defined in \cite{SM}, the friction coefficient produced by the cooling laser is given by
\begin{equation}
\beta=4\Omega^2\Gamma_1k_1\cos^2\theta(\frac{N_2}{N}-\frac{N_1}{N^2}),
\label{eq2}
\end{equation}
where $\Omega$ and $\Gamma_{1}$ are, respectively, the Rabi frequency and the linewidth regarding the excited state $|4P_{1/2}\rangle$, and $k_1$ is the wave number of the 397-nm cooling laser which irradiates the ions with respect to $z$-axis by $\theta$. $N$, $N_1$ and $N_2$ in Eq. (\ref{eq2}), as defined in \cite{SM}, are functions of the detunings $\Delta$ and $\delta$. In our trap system, the bath heating comes from thermal electron noise in the resistance of the electrodes in addition to some anomalous heating \cite{heating}, the former of which is usually called Johnson noise and the latter is still of unclear mechanism. For convenience of the treatment below, we resort all these noise to the bath heating strength $E_{e}$, and thus the ions' temperature $T$ after the Doppler cooling can be written as \cite{SM},
\begin{equation}
T=\frac{(\Delta-\delta)^2}{k_B N_h}+\frac{mE_{e}}{2k_B\beta},
\label{eq3}
\end{equation}
where the first term represents the heating effect due to photon scattering with the detuning-dependent $N_h$ defined in \cite{SM} and the second term is regarding the bath heating. Based on above equations, we find that the temperature can be fully controlled by the detuning $\Delta$ for a certain bath heating strength, as plotted in Inset (a) of Fig. \ref{Fig1}.

%On the other hand, the trapped $^{40}$Ca$^{+}$ ion in the Doppler cooling works as a three-level configuration coupled by two lasers (Inset (b) of Fig. \ref{Fig1}), where
%In what follows, the structural phase transition is witnessed in the SET by solely changing the temperature of the ions, which is actually controlled by the detuning of the 397-nm cooling laser.
%and the environmental heating effect is experimentally measured by observing the structural phase transition.
%For a certain bath, the temperature of the ions is mainly determined by $\beta$, which, in our case of $^{40}$Ca$^+$ ions, could be fully controlled by the detuning of the cool-ing laser [8]. As presented below, the structural phase transition is witnessed in the SET by solely changing the temperature of the ions, which is actually controlled by the detuning of the 397-nm cooling laser.

\begin{figure*}[htbp]
\centering
\includegraphics[width=18 cm,height=9 cm]{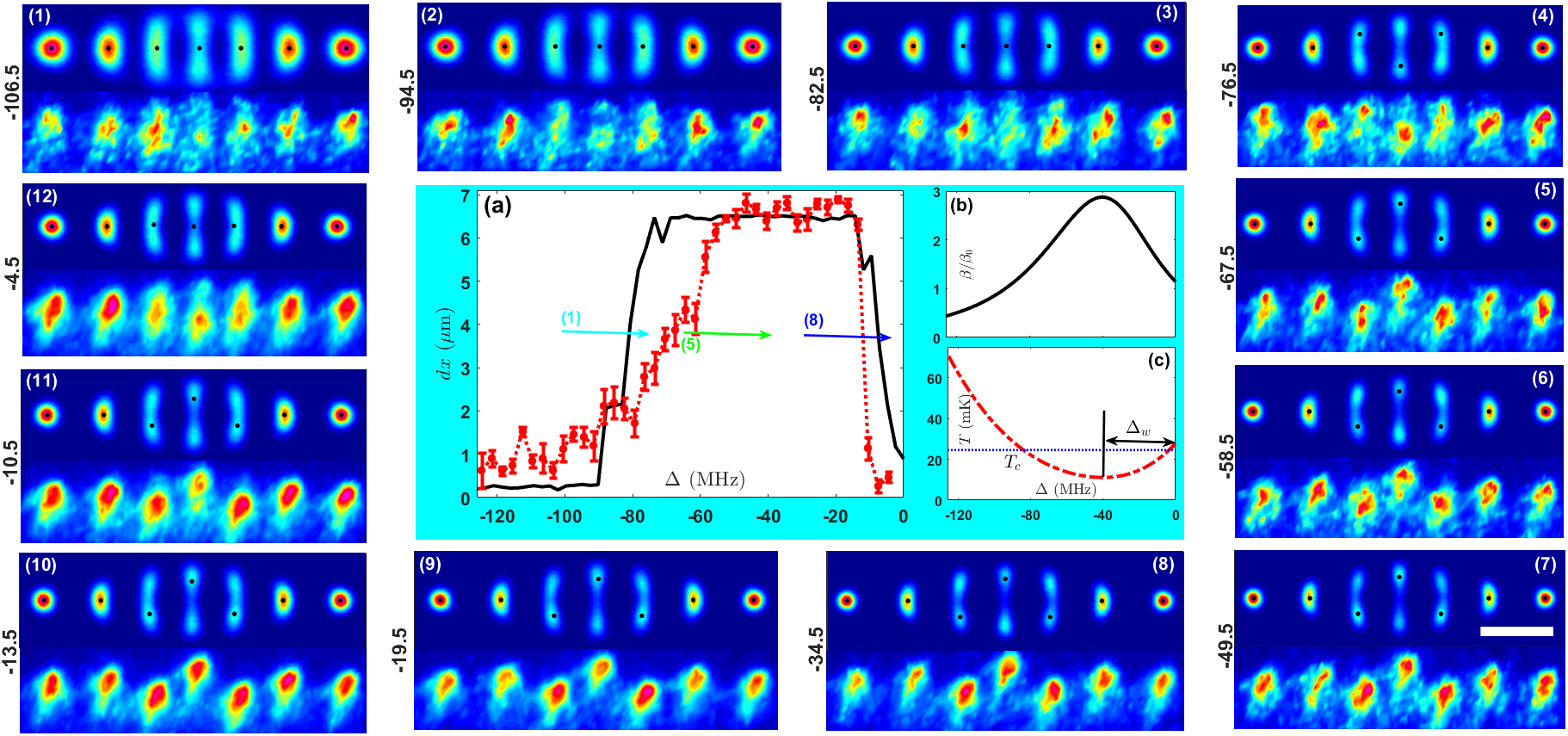}
\caption{SPT of seven $^{40}$Ca$^{+}$ ion crystals induced by the detuning $\Delta$ of the cooling laser. (a) Order parameter $dx$ in variation with the detuning, where the experimental observation (red dotted curve) is fitted by the numerical simulation (black solid curve) and the error bars of the experimental data (measured by 50 repetition) are determined by the mean square root. (b) Friction coefficient $\beta$ in units of $\beta_0=1\times 10^{-21}$ Ns$/$m as a function of the detuning $\Delta$. (c) Temperature $T$ of the ion crystals in $xz$ plane varying with the detuning $\Delta$, where $\Delta_w$ denotes the width of the detuning window, defined as the detuning difference from the point of zero friction coefficient to the point of the largest, and the blue dashed line indicates the critical temperature $T_c$.
%critical detuning -84(3) MHz, critical temperature 24(2) mK
The panels (1)-(12) present experimentally observed images (the lower of each panel) in comparison with numerically simulated results (the upper of each panel), where the number in the vertical axis of each panel is the detuning $\Delta$, corresponding to the phase transition steps labeled in (a). The horizontal direction means z-axis, the anisotropic parameter $\alpha=3.205$, and the bath heating intensity $E_e=13\mu_0$. A 20-$\mu$m scale bar is drawn in the panel (7) by considering the CCD resolution and 15 times magnification of the microscope objective before the CCD imaging. The scale bar applies to all the images.}
\label{Fig2}
\end{figure*}

Experimentally, to ensure a high-quality demonstration of this temperature-induced SPT, we have tried to stabilize the frequencies of both the 397-nm and 866-nm lasers by locking them to an optical cavity (linewidth of 3 MHz) made of a material with ultra-low expansion, and also employing the Pound-Drever-Hall technique. The incident directions of both the lasers are nearly parallel to the trap surface to minimize the scattering light due to the laser beams striking the surface of the trap, where the 866-nm laser with a small angle of $\pi/36$ with respect to the $xz$ plane provides a small component of cooling effect in the $y$-axis. The powers of 397-nm and 866-nm lasers are, respectively, $40~\mu$W and $250~\mu$W, which yield $\Omega/\bar{\Omega}=0.4$.
Besides, to minimize the influence from the rf heating, we have tried, by adjusting the compensation voltage $V_{AE}$, to keep the ion crystals initially close to the rf potential null, which is along the $y$-axis and above the trap surface by 910 $\mu$m. Moreover, our experiment gets started from the critical region of the SPT, which is obtained by checking a wide range of the parameters in advance by sweeping $\alpha$, as clarified later.
%To this end, we checked a wide range of the parameters in advance by sweeping $\alpha$ () and finally set our starting point close to the transition point $\alpha_{c}=$3.2785.

In the experiment, we confine seven ions in the surface-electrode trap by Doppler cooling, and then exactly raise the trapping potential to $\alpha=$3.205, which is accomplished by adiabatically increasing the voltage on the ME electrodes. We sweep the detuning $\Delta$ of the 397-nm laser from $-120$ MHz to $0$ MHz, during which the ion crystals experience a weak-strong-weak variation of the cooling efficiency, inducing a structural change from the linear to the zigzag and then back to the linear, as shown in Fig. \ref{Fig2}. The resolution blurring of the ions' images in our observation is due to both thermal effect of the ions resulted from the finite temperature (10 mK $\sim$ 65 mK) and the stray field noise. Nevertheless, considering the center of each ion, we could still identify the configurations of the ion crystals by means of a skillful treatment \cite{SM}.

Figure \ref{Fig2} indicates the observed configuration changes of the ion crystals in good agreement with the simulated results by the Langevin equation \cite{SM}. To characterize the structural changes, we employ the center-to-center distance $dx$ of two outermost ions in $x$-axis as the order parameter, which is very sensitive to the temperature change. As shown in Fig. \ref{Fig2}(a), we find an abrupt raising in the curves of $dx$ with respect to $\Delta$ around $\Delta_{c}=-84$ MHz and an abrupt falling around $\Delta_{c}=-5$ MHz, implying, respectively, the SPTs from the linear to the zigzag and back to the linear. Based on the analytical results in \cite{SM}, we may fully understand the phase transitions induced by the variations of the friction coefficient and the ions' temperature, see Fig. \ref{Fig2}(b,c). The critical temperature is $T_c=24(2)$ mK for both the abrupt raising and falling. The strongest friction of the cooling laser and the lowest temperature of the ion crystals appear at $\Delta=-40$ MHz, implying that the zigzag structure occurs at lower temperature than the linear chain. This is also reflected in the observation that the ions in the zigzag phase present clearer pictures even though those ions deviated from $z$-axis should suffer more serious rf heating. Considering all these associated factors, we deem that the laser cooling is dominant in this phase transition process.

\begin{figure*}[htbp]
\centering
\includegraphics[width=16 cm,height=9 cm]{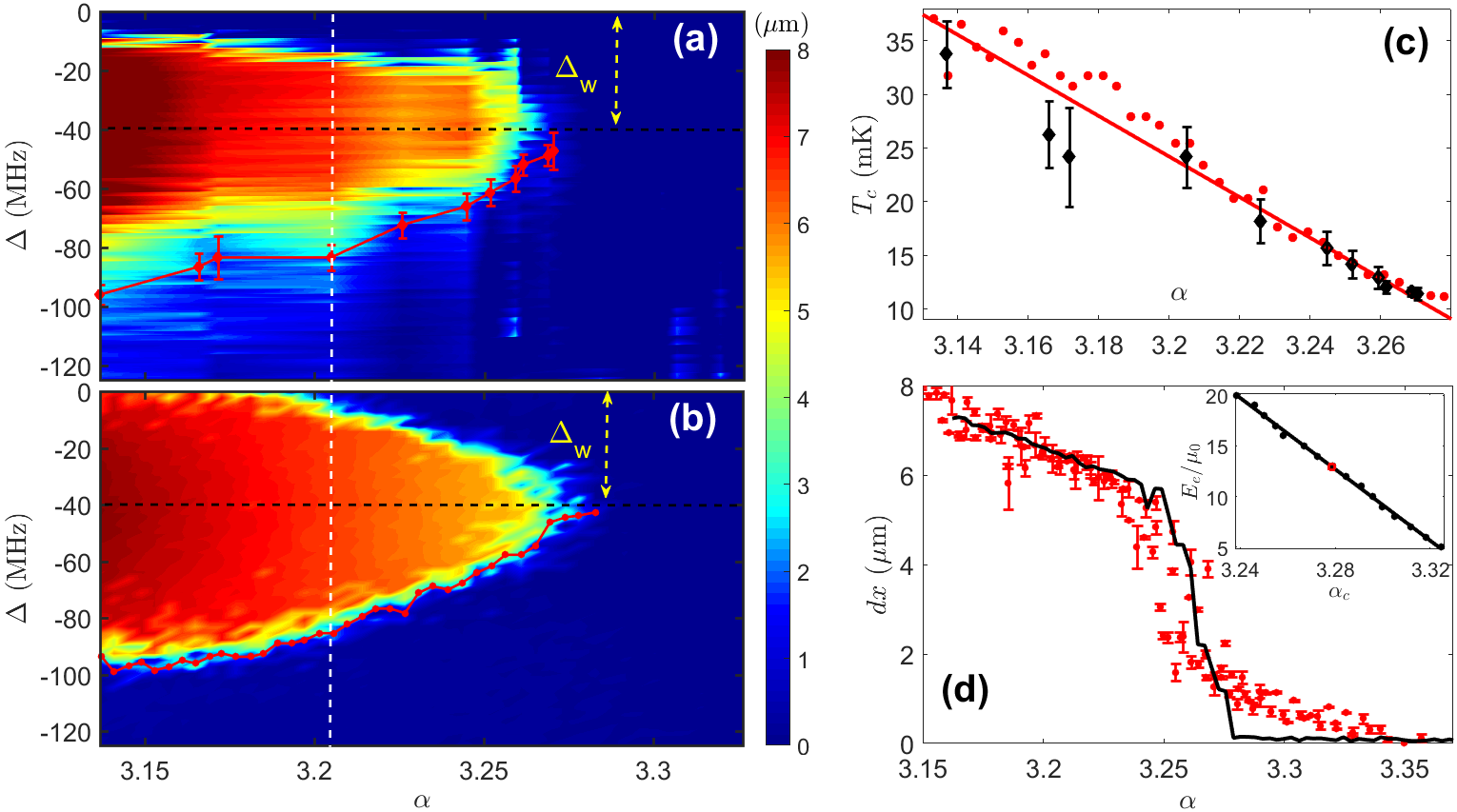}
\caption{(a, b) Phase diagrams of the SPT with respect to the detuning $\Delta$ and the anisotropic parameter $\alpha$, where the color bar indicates the values of $dx$. The vertical white dashed line labels the location of the observation demonstrated in Fig. \ref{Fig2}. The horizontal black dashed line indicates the position of the maximum order parameter $dx$. $\Delta_w$ denotes the width of the detuning window, as labeled in Fig. \ref{Fig2}(c). The red dots curves in (a) and (b) denote the critical detuning of SPT for different $\alpha$. (a) Experimental observation with each data measured by 50 repetition and (b) Numerical simulation with the bath heating intensity $E_e=13\mu_0$. (c) Critical temperature $T_{c}$ of the SPT vs $\alpha$, where red dots and black diamonds are obtained, respectively, by numerical simulation in (b) and experimental measurements in (a), and the red line is a linear fitting of the numerical results with $T_c=-189\alpha+629$. (d) Experimental observation of the SPT by sweeping $\alpha$ for $\Delta=-30$ MHz, where dots are experimental data measured by 20 repetition, and black solid curve is a numerical simulation with the transition point estimated as $\alpha_c=3.279(4)$. The error bars indicate standard deviation. Inset: Bath heating intensity $E_{e}$ as a function of $\alpha_c$ in the case of $\Delta=-30$ MHz, where dots are obtained by numerical simulation and the line denotes a linear fitting with $E_e/\mu_0=-176.3\alpha_c + 591$. The red circle indicates the transition point in our experiment.
%$\epsilon_1/\mu_0=176.3(4.5)$ and $\epsilon_2/\mu_0=591(15)$. the environmental heating intensity $\dot{E}_e=13.0(7)\mu_0$
}
\label{Fig3}
\end{figure*}

To give a more complete impression on this topic, we should fully characterize the scaling behavior at different transition points, which is also the prerequisite of understanding the present experiment. To this end, we have explored a wide range of $\Delta$ and $\alpha$ for the critical behavior of the phase transition (Fig. \ref{Fig3}). Although the strong bath heating makes the observation of the critical region blurry and deviated somewhat from numerical simulation, the experimentally obtained phase diagram could be in principle explained by numerical results from molecular dynamics method based on Langevin equation \cite{SM}. For example, we present particularly the comparison between experimental observation and theoretical simulation for the critical region of the abrupt raising (See the red dots in Fig. \ref{Fig3}(a,b)), which means that the anisotropic parameter $\alpha$ determines the critical detunings. In fact, $\alpha$ also determines the corresponding critical temperature, as plotted in Fig. \ref{Fig3}(c). As predicted in \cite{prl-105-265703}, the SPT occurs due to condensation of phonons into the soft mode - the lowest frequency collective motional mode of the ion crystals, which produces a linear scaling law between the critical temperature $T_c$ and the anisotropic parameter $\alpha$. Our observation in Fig. \ref{Fig3}(c) shows that the soft mode exists in the zigzag configuration, conforming to the previous prediction \cite{prl-105-265703,PRB77-064111}. As happened in finite temperature, the critical behavior is strongly sensitive to the thermal effects, which lead to anharmonic coupling between different phonon modes. The higher temperature of the ions brings about more complicated phonon coupling, implying more challenge to identify the phase transition critical behaviour. Thus, in Fig. \ref{Fig3}(c), there are larger errors and deviation observed at the higher temperature.

Besides, we should also measure the bath heating strength before implementing the phase transition. The theoretical study \cite{SM} has shown a linear relation of the phase transition point $\alpha_{c}$ with $E_e$, implying that $E_{e}$ can be measured by experimentally observing the position of critical point $\alpha_c$ (Inset of Fig. \ref{Fig3}(d)). In Fig. \ref{Fig3}(d), we choose $\Delta=-30$ MHz and find the transition point $\alpha_{c}=$3.279(4), which gives $E_e=13.0(8)\mu_0$. As such, temperature $T$ in our experiment can be indirectly detected based on Eq. (\ref{eq3}), for which we have proposed an effective method to obtain $\Omega$ by measuring $\Delta_{w}$ in a linear relation \cite{SM}. Due to short lifetime of the $|4P_{1/2}\rangle$, there was no experimental means to directly detect the Rabi frequency regarding the cooling laser. But with our method, we obtain $\Omega=78(17)$ MHz by measuring $\Delta_{w}=40(6)$ MHz in Fig. \ref{Fig3}, and thus $T$ is obtained.

As final remarks, we mention following important points. The discrepancy between the experimental values and the simulated results indicates the imperfection in our operations with respect to the ideal consideration. The main source of errors is the heating of the
thermal bath of the ions, which keeps bringing in thermal noise and leads to measurement imprecision. Besides, frequency instability of the lasers also contributes $4\%$ error, but the detrimental influence from the rf heating is negligible. In above figures, part of the influence due to bath heating on the measurement is resorted to the error bars as statistical errors, and others yield the experimental values deviated from the simulated results.
%To monitor the temperature variation of the ion crystals during our operations, we proposed an effective method to obtain $\Omega$ by measuring $\Delta_{w}$.
Moreover, to avoid the complexity due to the dark resonance in our case \cite{NJP17-045004,SM}, we have to keep the 866-nm laser slightly blue-detuned throughout the experiment, which is practically accomplished by sweeping the 866-nm laser to the point that the cooling starts weakening. This makes sure that the temperature of the ions is strictly controlled by tuning the frequency of the 397-nm laser.

%Furthermore, with respect to the cusp-like phase transition in the thermodynamic limit, the system with seven ions only shows the abrupt raising with definite slope around the transition point of the phase transition for the experimental observation in Fig. \ref{Fig2}(a) and Fig. \ref{Fig3}(d). The cusps appeared in the numerical simulation are due to discrete conditions in our data processing \cite{SM}.

In conclusion, we have scrutinized the thermal fluctuation effect in a microscopic thermodynamic process by witnessing, for the first time, a temperature-induced phase transition at atomic level, i.e., a linear-zigzag pattern change in ion crystals under controllable influence of thermal fluctuation. Different from the temperature-relevant cloud-order phase transitions in ion traps as observed previously in \cite{order1,order2,order3}, the trapped ions in our case have just changed the crystalline conformation throughout the experimental process. More importantly, our observation of bidirectional changes between the linear and zigzag structures undoubtedly indicates that thermal fluctuation could play an essential role, even a positive role, in microscopic thermodynamics, which could never been understood in the conventional thermodynamics regarding the macroscopic systems. As such, the trapped-ion system in conformation variation provides a good research platform for reconsidering thermodynamic quantities and thermodynamic processes subject to the fluctuation theory \cite{flu1,flu2}. On the other hand, with further cooling of the ion crystals down to zero temperature, the SPT under consideration can straightforwardly be mapped into a quantum phase transition of Ising model subject to a transverse field \cite{Shimshoni,Schmied,Bermudez}. In this sense, a phase transition from the classical to the quantum would be able to happen \cite{RMP84-1655,PRB89-214408}, and thus more observations of fascinating quantum behavior under quantum fluctuation would be expected.

This work was supported by National Key R$\&$D Program of China under grant No. 2017YFA0304503, by National Natural Science Foundation of China under Grant Nos. 11835011, 11804375, 11734018, 11674360 and 91421111, and by the Strategic Priority Research Program of the Chinese Academy of Sciences under Grant No. XDB21010100.

\end{document}